\documentstyle[12pt,a4wide]{article}
\begin{document}
\title{\bf Electrostatic self-energy in \\
static black holes with spherical symmetry}
\author{B. Linet \thanks{E-mail: linet@celfi.phys.univ-tours.fr} \\ 
\small Laboratoire de Math\'ematiques et Physique Th\'eorique \\
\small CNRS/UPRES-A 6083, Universit\'e Fran\c{c}ois Rabelais \\
\small Parc de Grandmont 37200 TOURS, France}
\date{}
\maketitle
\thispagestyle{empty}

\begin{abstract}

We determine the expression of the electrostatic self-energy for
a point charge in the static black holes with spherical symmetry 
having suitable properties.

\end{abstract} 

\section{Introduction}

A renewed interest in the electrostatic self-energy for a point charge
in black holes has appeared in studies of the upper bound on the entropy 
for a charged object \cite{bek1,hod,lin1}. For a neutral object, the original 
method required the validity of the 
generalized second law of thermodynamics for the Schwarzschild black 
hole \cite{bek2}. In the charged case, it is moreover
essential to take into account the electrostatic self-energy on the horizon 
of the charged object within the method. For the Schwarzschild
black hole, the expression of the self-force acting on a point charge at
rest has been determined for a long time \cite{smi,zel,lea1}
because the electrostatic potential was known in closed form 
\cite{cop,lin2} and so the entropy bound has been found by this way. 
However, we have recently taken up the question
of how derive the entropy bound for a charged object by employing 
thermodynamics of any static black holes with spherical symmetry
\cite{lin3}. A
crucial step in this case is again the determination of electrostatic 
self-energy. Unfortunately, our expression of the electrostatic 
self-energy was mainly conjectured from some strong indications but the 
precise assumptions on the black holes ensuring this result was not
indicated.

The purpose of this work is to determine of the expression of the electrostatic
self-energy for a point charge in the static black holes with spherical 
symmetry verifying the following general assumptions.
\begin{enumerate}
\item The coordinate system in which the metric is manisfestly static must 
describe entirely the spacetime outside the horizon of the black hole.
However, it is not necessarily asymptotically Minkowskian.
\item The surface gravity of the horizon of the black hole is
different from zero.
\item When the black hole is charged, then the electrostatic potential
generated by this charge tends to zero at the spatial infinity.
\item The no hair conjecture for the electric field
holds for the black hole, i.e. there is no black hole with electric 
multipole moments, except with the monopole.
\end{enumerate}
In fact, the third and fourth assumptions induce the same properties
on the test electric field in the background geometry of the given
black hole. In our procedure, these assumptions 
have to be directly verified on the test electric multipole fields.
 
The plan of the work is as follows. We recall in section 2 some useful
formulas about the static black holes with spherical symmetry. 
In section 3 we discuss the electrostatic equation in the black holes 
verifying the above mentioned assumptions. We calculate in section 4 
the electrostatic self-energy on the horizon. We add some concluding 
remarks in section 5.

\section{Preliminaries}

If we suppose that the spacetime describing the static black hole with
spherical symmetry is such that the area of the spheres 
increases with the radial coordinate, then its metric can be written as
\begin{equation}\label{0}
ds^2=g_{tt}(R)dt^2+g_{RR}(R)dR^2+R^2(d\theta^2+\sin^2\theta d\varphi^2)
\end{equation}
in a coordinate system $(t,R,\theta ,\varphi )$. 
The component $g_{tt}$ vanishes at the horizon located at $R=R_H$
with $R_H>0$. By virtue of the first assumption,
metric (\ref{0}) is well defined for $R_H<R<\infty$. In this domain,
we have $g_{tt}<0$ and $g_{RR}>0$ but for a black hole $g_{RR}$ becomes 
singular as $R\rightarrow R_H$.
The surface gravity of the horizon $\kappa$ has the general expression
\begin{equation}\label{2s}
\kappa =\frac{\partial_Rg_{tt}}{2\sqrt{-g_{tt}g_{RR}}}\bigg|_{R=R_H} .
\end{equation}

In our procedure, we have needed to write down metric (\ref{0}) in 
isotropic form. To do this, we perform a change of radial coordinate 
$R(r)$ defined by the differential equation
\begin{equation}\label{i}
\sqrt{g_{RR}(R)}\frac{dR}{dr}=\frac{R}{r} .
\end{equation}
It follows from (\ref{i}) that the radial coordinate $r$ is determined up
to an arbitrary factor but it is well defined outside the horizon.
We denote $r_H$ the corresponding value of $R_H$ and we remark that
$r_H\geq 0$. In the coordinate system $(t,r,\theta ,\varphi )$, 
metric (\ref{0}) of the black hole can be now written as
\begin{equation}\label{1}
ds^2=-N^2(r)dt^2+B^2(r)\left( dr^2+r^2d\theta^2+r^2\sin^2\theta
d\varphi^2\right)
\end{equation}
in which the horizon is located at $r=r_H$. In the domain $r_H<r<\infty$, 
metric (\ref{1}) is well defined.   
Since the hypersurface $r=r_H$ is a horizon, we have $N(r_H)=0$. 

The second assumption, $\kappa \not= 0$,
implies that $\partial_rN(r_H)\not= 0$ and $\lim_{r\rightarrow r_H}B(r)$
finite. This excludes $r_H=0$ and $B(r_H)$ being finite, 
we have $B(r_H)\not= 0$ since the area of the black hole does not vanish.
Hence from the general expression (\ref{2s}), we get 
\begin{equation}\label{2}
\kappa =\frac{N'(r_H)}{B(r_H)}
\end{equation}
where the prime signifies the differentiation with respect to $r$.
Taking into account (\ref{2}), we see immediately that
\begin{equation}\label{3}
N(r)\sim \kappa B(r_H)(r-r_H) \quad {\rm as}\quad r\rightarrow r_H .
\end{equation}
with $\kappa B(r_H)\not= 0$.

\section{Electrostatic potential}

In the static case,
the Maxwell equations in background ({\ref{1}) yields the following equation
for the electrostatic potential $A_t$
\begin{equation}\label{4}
\frac{1}{\sqrt{-g}}\partial_i\left( \sqrt{-g}g^{tt}g^{ij}\partial_jA_t\right)
=4\pi J^t
\end{equation}
where $J^t$ is the charge density. For a point charge $e$ at $r=r_0$, 
$\theta =\theta_0$ and $\varphi =\varphi_0$, 
it has the expression
\begin{equation}\label{5}
J^t(r,\theta ,\varphi )=
\frac{e}{\sqrt{-g}}\delta (r-r_0)\delta (\theta -\theta_0)
\delta (\varphi -\varphi_0) .
\end{equation}
Without loss of generality, we may write down the electrostatic equation 
(\ref{4}) for $\theta_0=0$ in the form
\begin{equation}\label{8}
\triangle A_t+\frac{N}{B}\left( \frac{B}{N}\right) '
\partial_rA_t =-4\pi e \frac{N}{r^2B}\delta (r-r_0)
\delta (\cos \theta -1)
\end{equation}
where $\triangle$ is the usual Laplacian operator. 

The electric field derived from the electrostatic potential, 
solution to equation (\ref{8}), should 
be well behaved at the horizon and at the spatial infinity of the black hole.
Now we require the absence of charge 
inside the horizon, therefore there is no electric flux through the horizon. 
Consequently, the Gauss theorem gives
\begin{equation}\label{7}
\int_{r_1}\frac{\sqrt{-g}}{\sin \theta}g^{tt}g^{rr}\partial_rA_t
d\theta d\varphi =\left\{ \begin{array}{ll}
4\pi e & r_1>r_0 \\
0 & r_H<r_1<r_0 . \end{array}\right. 
\end{equation}

We can expand $A_t$ in spherical harmonics. In the axially symmetric case, 
we put
\begin{equation}\label{9}
A_t(r,\theta )=\sum_{l=0}^{\infty}R_l(r,r_0)P_l(\cos \theta )
\end{equation}
where the function $R_l$ obeys the differential equation
\begin{equation}\label{10}
R''_l+\left( \frac{2}{r}+\frac{B'}{B}-\frac{N'}{N}\right) R'_l 
-\frac{l(l+1)}{r^2}R_l=-e(2l+1)\frac{N}{r^2B}\delta (r-r_0) .
\end{equation}
The problem to determine $R_l$ reduces to define two linearly
independent solutions $g_l$ and $f_l$ of the homogeneous differential
equation (\ref{10}) with appropriate boundary conditions.

In the case $l=0$ of equation (\ref{10}), an integration leads to
$$
\partial_rR_0(r)={\rm const.}\times \frac{N(r)}{r^2B(r)} .
$$
By virtue of our third assumption, the test electric field in background
(\ref{1}) generated by a charge inside the horizon has an electrostatic
potential vanishing at the spatial infinity and regular at the 
horizon. So, we can put
\begin{equation}\label{11o}
g_0(r)=1 \quad {\rm and} \quad f_0(r)=a(r) \quad {\rm with} \quad
a(r)=\int_{r}^{\infty}\frac{N(r)dr}{r^2B(r)} .
\end{equation}
According to the Gauss theorem (\ref{7}), $a$ is the electrotatic potential
generated by a unit charge inside the horizon. It is finite at
$r=r_H$ and we set $a(r_H)=a_H$.

In the case $l\not= 0$ of equation (\ref{10}), the point $r=r_H$
is a singularity of the homogeneous differential equation. From (\ref{3}),
we see that 
$$
\frac{N'}{N}\sim \frac{1}{r-r_H} \quad {\rm as}\quad r\rightarrow r_H
$$
and consequently the point $r=r_H$ is a singular point of regular type of the 
differential
equation (\ref{10}). The roots of the indicial equation relative to
this point are 0 and 2. Thus, there exists a regular solution at $r=r_H$,
noted $g_l$, such that
\begin{equation}\label{11l}
g_l(r)\sim (r-r_H)^2 \quad {\rm as} \quad r\rightarrow r_H 
\end{equation}
and so the corresponding electric field is well behaved on the horizon. 
The solution $g_l$ cannot regular as 
$r\rightarrow \infty$ because the test electric field would be well behaved 
at the spatial infinity and this fact would be in contradiction with the fourth
assumption which demands the nonexistence of black hole with multipole
electric moments, except with the monopole. 
Consequently, the solution $g_l$ is singular as $r\rightarrow \infty$.
We call $f_l$ the regular solution as $r\rightarrow \infty$. By using the
same argument, we find that the solution $f_l$ is singular at $r=r_H$. 

Therefore, the electrostatic potential (\ref{9}) having the adequate
boundary conditions can be written down in the form
\begin{equation}\label{12}
A_t(r,\theta )=\left\{ \begin{array}{ll}
\displaystyle ea_H+\sum_{l=1}^{\infty}eC_lg_l(r)f_l(r_0)P_l(\cos \theta ) & r_H<r<r_0 \\
\displaystyle ea(r)+\sum_{l=1}^{\infty}eC_lg_l(r_0)f_l(r)P_l(\cos \theta ) & r>r_0
\end{array}\right.
\end{equation}
where the constants $C_l$ are uniquely determined by equation (\ref{10}). 

We now return to the partial differential equation (\ref{8}). As the second
partial derivatives in this operator take the form of the usual Laplacian,
the behaviour of $A_t$ at the neighbourhood of the point $r=r_0$ and 
$\theta =0$ is given by
\begin{equation}\label{14}
A_t(r,\theta )\sim \frac{N(r_0)}{B(r_0)}\times \frac{e}
{\sqrt{r^2-2rr_0\cos \theta +r_{0}^{2}}} . 
\end{equation}

\section{Electrostatic self-energy}

We consider the electrostatic energy associated with the Killing vector 
$\partial_t$ of metric (\ref{1}). The Coulombian part (\ref{14}) of 
the electrostatic potential $A_t$ does not yield an electrostatic self-force.
As shown in the previous works \cite{smi,zel,lea1}, 
the regular part of $A_t$ at $r=r_0$ and 
$\theta =0$ enables us to define the electrostatic self-energy 
$W_{self}(r_0)$ by the following limit process
\begin{equation}\label{15}
\frac{e}{2}\left( A_t(r,\theta )-\frac{N(r_0)}{B(r_0)}\times
\frac{e}{\sqrt{r^2-2rr_0\cos \theta +r_{0}^{2}}}\right) \rightarrow
W_{self}(r_0)\quad {\rm as}\; r\rightarrow r_0\; \theta \rightarrow 0 .
\end{equation}
However, since we do not know in general the expression of $A_t$ in closed 
form, it is difficult to evaluate $W_{self}(r_0)$ by using (\ref{15}).

In order to calculate (\ref{15}), we consider the explicit function $V_C$ 
introduced by Copson in the Schwarzschild metric characterized by the mass 
$M$ \cite{cop}. It is the solution
in the Hadamard sense of the electrostatic equation in isotropic coordinates,
i.e. equation (\ref{8}) with the coefficient
$$
\frac{N^S(r)}{B^S(r)}=\frac{1-M/2r}{\left( 1+M/2r\right)^3} .
$$
We choose $M=2r_H$ so that the horizon of the Schwarzschild black hole
in isotropic coordinates coincides with $r_H$. 
This function $V_C$ has also the same behaviour (\ref{14}) in a 
neighbourhood of the point $r=r_0$ and $\theta =0$ which is given by
\begin{equation}\label{16}
V_C(r,\theta )\sim \frac{\left( 1-M/2r_0\right)}
{\left( 1+M/2r_0\right)^3}\times \frac{e}
{\sqrt{r^2-2rr_0\cos \theta +r_{0}^{2}}} .
\end{equation}
We are now in a position to find a new limit process, instead of (\ref{15}), 
by replacing   
$e/\sqrt{r^2-2rr_0\cos \theta +r_{0}^{2}}$ by $V_C(r,\theta )$ with
the appropriate factor which takes into account (\ref{16}). 
We thus have 
\begin{equation}\label{15s}
\frac{e}{2}\left( A_t(r,\theta )-\frac{N(r_0)}{B(r_0)}\times
\frac{\left( 1+M/2r_0\right)^3}{\left( 1-M/2r_0\right)}
V_C(r,\theta )\right) \rightarrow W_{self}(r_0) \quad {\rm as}\; r\rightarrow
r_0\; \theta \rightarrow 0 .
\end{equation} 

The solutions $g_{l}^{S}$ and $f_{l}^{S}$ are known in function of
the radial coordinate $R$ of the Schwarzschild metric,
likewise the constant $C_{l}^{S}$ \cite{coh}. Of course, the solutions
$g_{l}^{S}$ and $f_{l}^{S}$ satisfy the desired boundary conditions
because the no hair theorem for the Schwarzschild black hole 
has been proved \cite{isr}. The analysis of the explicit expression 
of the Copson solution $V_C$ with the aid of the Gauss
theorem (\ref{7}) shows that it describes furthermore a charge 
$-eM/r_0(1+M/2r_0)^2$ inside the horizon \cite{lin2}. In the multipole 
expansion (\ref{9}) of $V_C$, the monopole term must take into account 
this fact. We have thereby
\begin{equation}\label{12s}
V_C(r,\theta )=\left\{ \begin{array}{ll}
\displaystyle \frac{e}{r_0\left( 1+M/2r_0\right)^2}\left( 1-\frac{M}
{r\left( 1+M/2r\right)^2}\right) & \\ 
 & \\
\quad \quad \displaystyle +\sum_{l=1}^{\infty}
eC_{l}^{S}g_{l}^{S}(r)f_{l}^{S}(r_0)P_l(\cos \theta ) & r_H<r<r_0 \\
\displaystyle \frac{e}{r\left( 1+M/2r\right)^2}\left( 1-\frac{M}
{r_0\left( 1+M/2r_0\right)^2}\right) & \\ 
& \\
\quad \quad \displaystyle +\sum_{l=1}^{\infty}
eC_{l}^{S}g_{l}^{S}(r_0)f_{l}^{S}(r)P_l(\cos \theta ) & r>r_0 .
\end{array}\right. 
\end{equation}

We now insert the multipole expansions (\ref{12}) and (\ref{12s}) into 
the left term of formula (\ref{15s}). We simply set $r=r_0$ and
$\theta =0$ in the resulting infinite series to evaluate $W_{self}(r_0)$.
Now we are only interested in the determination of the electrostatic 
self-energy on the horizon, denoted $W_{self}$, therefore we take the 
limit $r_0\rightarrow r_H$ in this infinite series.  
The important point in the limit process is that
\begin{equation}\label{3b}
\lim_{r_0\rightarrow r_H}
\frac{N(r_0)}{r_H\left(1-M/2r_0\right)B(r_H)} =\kappa 
\end{equation}
which can be derived from (\ref{3}).
Now each term $l=1,2,\ldots$ of the infinite series contains $g_l(r_H)$ or
$g_{l}^{S}(r_H)$ which vanish according to (\ref{11l}), therefore it remains
only to study in (\ref{15s}) the monople terms of the multipole expansions
(\ref{12}) and (\ref{12s}). Due to (\ref{3b}), we obtain finally
\begin{equation}\label{21}
W_{self}=\frac{e^2}{2}(a_H-\kappa ) .
\end{equation}

For another Killing vector $\partial_{\overline{t}}$ resulting from the
rescaling of the time coordinate
$\overline{t}=\lambda^2t$, we easily see that $\overline{W}_{self}$
is given $W_{self}/\lambda$ since $\overline{a}_H=a_H/\lambda$ and
$\overline{\kappa}=\kappa /\lambda$. If metric (\ref{1}) is
asymptotically Minkowskian, then we can normalize $\partial_t$. 

\section{Conclusion}

For black holes verifying the prescribed assumptions,
we have provided a method for the calculation of the electrostatic
self-energy on the horizon, fortunately without having to know
the expression of the electrostatic potential in closed form. 
The key point is the limit process (\ref{15s}).  
We emphasize that only the knowledge of the monopole in the multipole 
expansion of the electrostatic potential is required to calculate the
electrostatic self-energy. 
From (\ref{21}), we can immediately show that the
electrostatic self-energy at the position $r=r_0$ is given by
\begin{equation}\label{20}
W_{self}(r_0)=\frac{1}{2}e^2s[a(r_0)]^2 \quad {\rm with}\quad s=\frac{1}{a_H}
\left( 1-\frac{\kappa}{a_H}\right)
\end{equation}
which agrees with our previous conjecture \cite{lin3}. Of course, this
formula is independent on the choice of the radial coordinate.    

In the present proof, we have required that the surface gravity of the horizon
$\kappa$ is different from zero. It is probably not necessary
for all extreme black holes. Indeed,
formula (\ref{21}) or (\ref{20}) gives the electrostatic self-energy
in the extreme Reissner-Nordstr\"{o}m black hole for which $\kappa =0$.
We can directly see it from the expression of the electrostatic 
potential in the Reissner-Nordstr\"{o}m black hole which is known
in closed form \cite{lea2}.

\end{document}